# Entertaining the Possibility of RT Superconductivity in LK-99

Itai Panas,

Department of Chemistry and Chemical Engineering,

Chalmers University of Technology, Gothenburg, SE 41296, Sweden.

**Abstract**

An intuitive chemical perspective on the LK-99 material is outlined and supported by DFT calculations. A hidden "flat" Lead band that exhibits instability toward charge density wave formation is exposed. Electron transfer between Lead CDW/conduction bands and $Cu3d^9/Cu3d^{10}$ impurity states is suggested to embody the observed phenomenology reminiscent of room temperature superconductivity. The inter-system electronic instability is reflected in a chemical instability involving $2Pb_{10-x}Cu_x(PO_4)_6O \Rightarrow Pb_{20-2x}Cu_{2x}(PO_4)_{12}(O_2)$. Implications for tuning posttreatments as well as handling of the LK-99 material emerge.

Recent reports of room temperature superconductivity in the compound LK-99 of nominal composition $Pb_{10-x}Cu_x(PO_4)_6O$ [1,2] have inspired us to perform exploratory DFT calculations over the past two weeks. And this, in a complementary direction to that of recent publications in Arxiv, see e.g. [3,4]. Our independent findings have led us to encourage and support further experiments on the compound LK-99 including soft reduction in conjunction with low-temperature annealing and corresponding tuning of the copper content to expose the intriguing yet elementary interplay between lattice and electrons, and between Lead and copper, inherent in this system. The conclusion is based on the following:

1. Consider first the nominal $Pb_{10}(PO_4)_6O$. This is a layered system, with Lead planes defined by triangles in the ab-plane and Lead chains along the c-direction. Phosphate groups hold the structure together, the phosphorous atoms are co-planar with the corresponding Lead triangles. Exercising the electron counting for Lead and phosphates, assuming $Pb^{2+}$ and $PO_4^{3-}$, leaves us with two "extra" electrons. These are absorbed by the nominal "extra" oxygen. Removing the "extra" oxygen from the structure, the "extra" electrons end up in the 6p-band of the in-plane Lead ions hybridized to form 3-centred bonds pointing into the center of the triangles. In as much as the unit cell is composed of two planes, formally, each layer should carry one of the two "extra" electrons. However, the system is found to produce a charge density wave CDW where every second plane carries two electrons and every second none, see Figure 1a. Therefore, the Lead triangles are found to contract or expand depending on whether reduced ($R_{Pb-Pb}\approx3.8$ Å) or oxidized ($R_{Pb-Pb}\approx4.3$-$4.4$ Å). This CDW can be made to dissolve by performing the excitation from the diamagnetic to the ferromagnetic state, excitation energy of $\approx0.13$ eV, see Figure 1b. In the ferromagnetic state, each triangular plane gets to carry one "extra" electron and consequently the Lead triangles come out equivalent ($R_{Pb-Pb}\approx4.1$ Å). All this has remarkably small effects on the overall unit cell dimensions. On introducing the "extra" oxygen, it is found to enter one of the triangular sites thereby locking the "inherited" CDW structure.

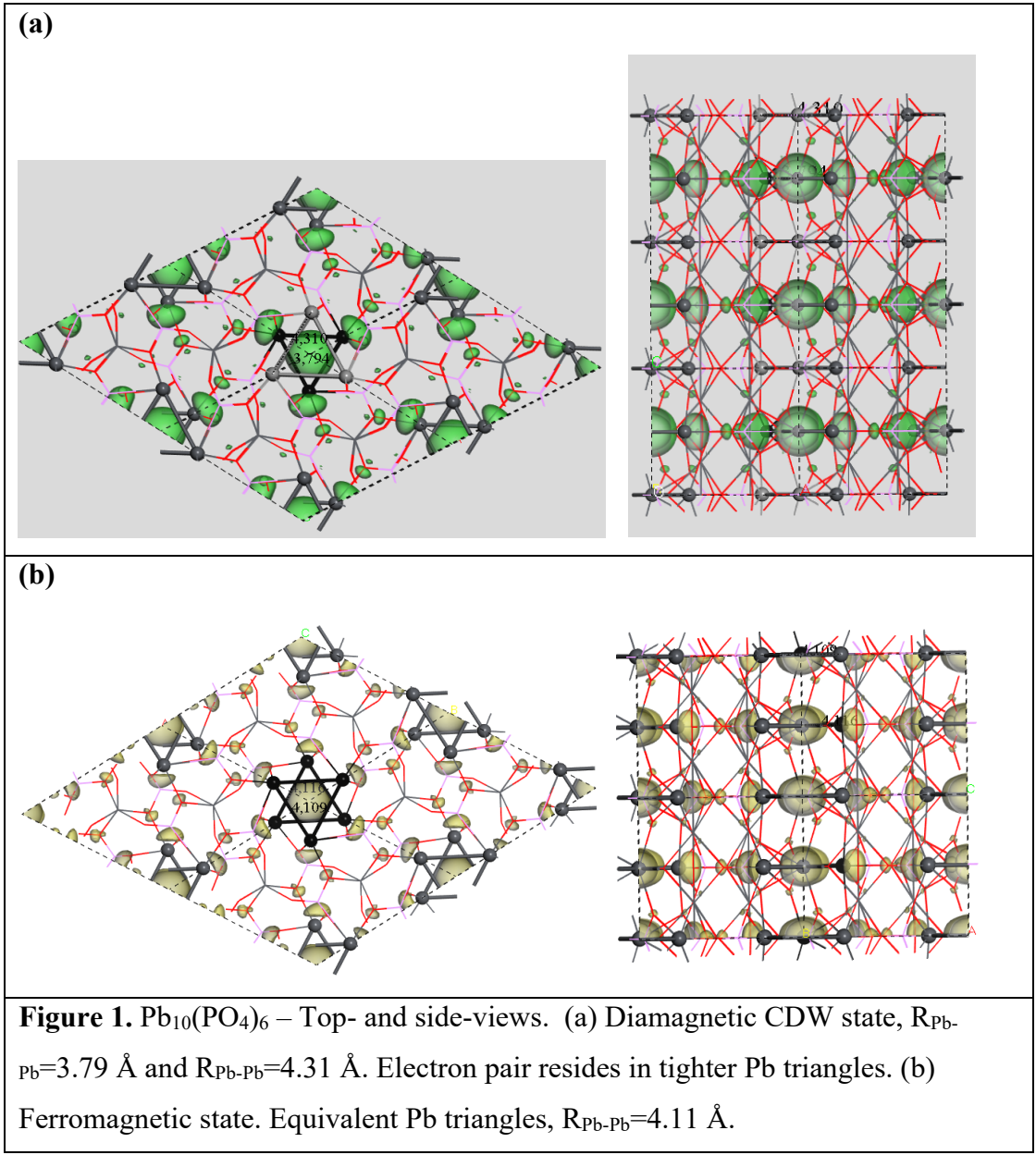

**Figure 1.** $Pb_{10}(PO_4)_6$ – Top- and side-views. (a) Diamagnetic CDW state, $R_{Pb-Pb}$=3.79 Å and $R_{Pb-Pb}$=4.31 Å. Electron pair resides in tighter Pb triangles. (b) Ferromagnetic state. Equivalent Pb triangles, $R_{Pb-Pb}$=4.11 Å.

2. Returning to $Pb_{10}(PO_4)_6O$. On comparing the DFT structure to those obtained from crystallography, the most outstanding deviation is the position of the "extra" oxygen, which DFT places in the center of the Lead triangles, while XRD has the oxygens displaced on either side of the planes by $\approx \pm 0.85$Å, (*cf.* $Pb_{10}(PO_4)_6O$-KB in reference [5]). This inconsistency between XRD and DFT may be resolved by doubling the unit cell along the c-direction and associating a peroxide ion $O_2^{2-}$ to every four Lead planes. Because of the unit cell doubling, now four "extra" electrons must be accommodated, two of which are accounted for by the $O_2^{2-}$. And indeed, periodic contractions and expansions of triangles along the c-direction is found, i.e., contraction of the triangle owing to $O_2^{2-}$ in one plane is followed by expansion of oxidized triangles in the adjacent plane, contraction of reduced triangle accommodating the second pair of "extra" electrons subsequently, expansion of the third oxidized plane, and finally contraction of Lead triangle around a second peroxide ion, see Figure 2. The obtained $R_{O-O} \approx 1.53$ Å corresponds to oxygens displaced on either side of every 4$^{th}$ Lead plane by $\approx \pm 0.76$ Å is in fair agreement with the reinterpreted XRD data.

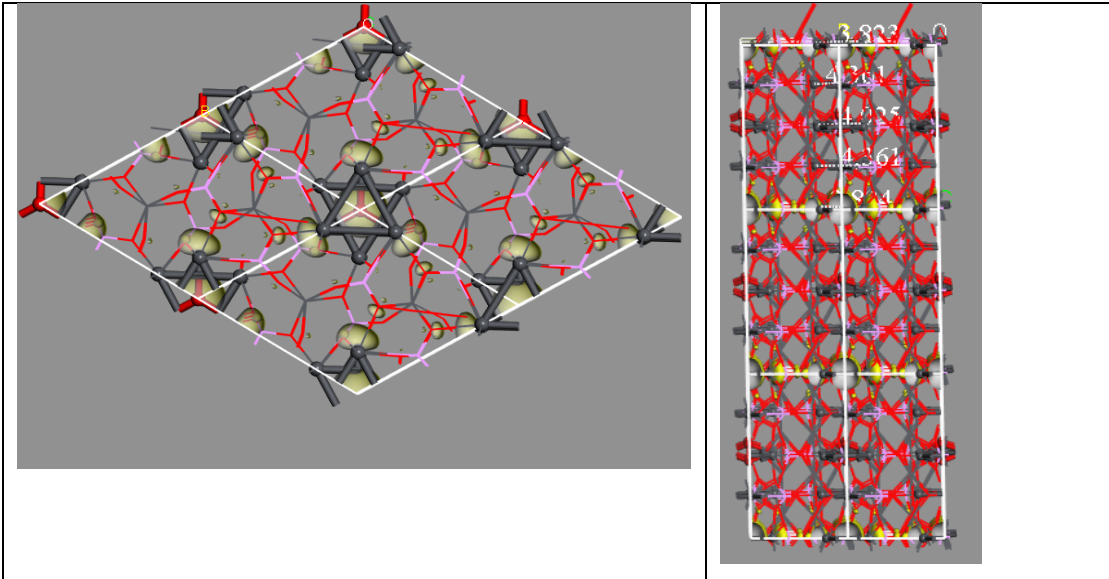

**Figure 2.** $Pb_{20}(O_2)(PO_4)_{12}$ –Top- and side-views of CDW. Of the four "extra" electrons, two contribute to peroxide $R_{Pb-Pb}=4.02$ Å. The other two "extra" electrons reside in next-nearest neighboring planes $R_{Pb-Pb}=3.82$ Å. For nearest neighbor planes to the peroxide plane, $R_{Pb-Pb}=4.36$ Å.

3. Now we turn to LK-99, which owing to (2) is taken here to comprise $Pb_{20-2x}Cu_{2x}(PO_4)_{12}(O_2)$, $0.9<x<1.1$. To understand the impact of nominal substitution of $Pb^{2+}$ for $Cu^{2+}$, we first remind ourselves of the two "extra" electrons that reside in three-fold sites of every four Lead planes. Secondly, we make the general observation that the ionic radius of $Pb^{2+}$ is significantly larger than that of $Cu^{2+}$. This renders the stability of $Cu^{2+}$ compromised when being accommodated in the $Pb^{2+}$ sublattice. Therefore, the resulting weaker crystal field rather favors $Cu^+$. Consequently, on optimizing the Cu position in $Pb_{19}Cu(PO_4)_{12}(O_2)$, it is found to take a Cu-O distance of 2.04 Å which is rather long as compared to 1.95 Å for $3d^9$ $Cu^{2+}$ in for example the hole doped cuprate superconductors. No spin density is found in this case on Cu, consistent with the $3d^{10}$ electron configuration, see Figure 3. As only one of the two "extra" electrons remains in the triangular Lead site, correspondingly, an expansion of the Lead triangle from 3.83 Å to $R_{Pb-Pb} \approx 4.2$ Å is observed reflecting the reduced screening in the site. Thus, on replacing a second $Pb^{2+}$ by $Cu^{2+}$ the second "extra" electron in understood to be absorbed by formation of a second $3d^{10}$ $Cu^+$. Chemically, this oxidation of the Lead sub-lattice is understood to further enhance the preference for peroxide formation. This follows if considering that charge neutrality requires that $2Cu^{2+} \leftrightarrow 2O^{2-}$ while $2Cu^+ \leftrightarrow [2O^-] \leftrightarrow O_2^{2-}$ + vacancy.

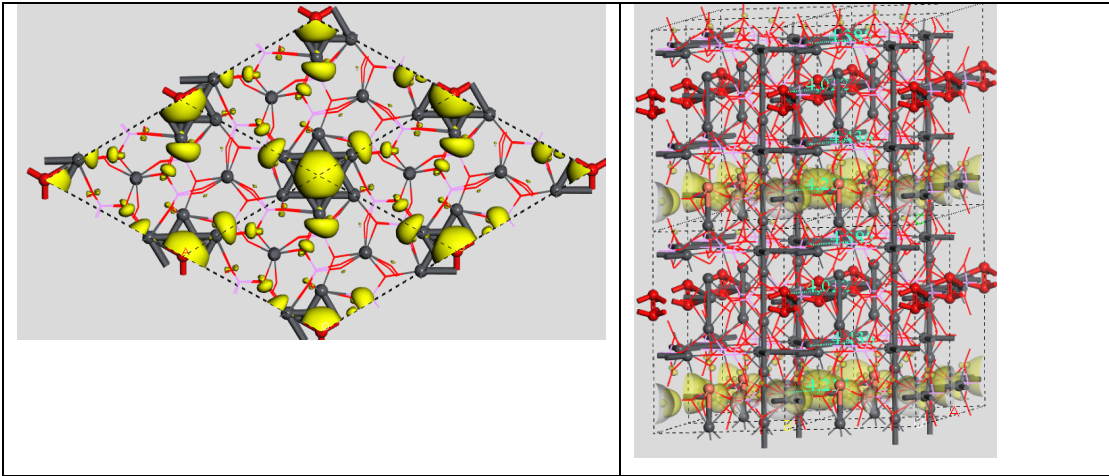

**Figure 3.** Top- and side-views. Spin density in $Pb_{19}Cu(PO_4)_6(O_2)$ - No spin density on Cu corresponds to $3d^{10}Cu^+$. The residual "extra" electron is situated in CDW associated plane, $R_{Pb-Pb}$=4.23 Å. Peroxide associated plane, $R_{Pb-Pb}$=4.02 Å. The $R_{Pb-Pb}$ distances in the intervening unscreened planes are 4.39 Å and 4.43 Å.

4. Indeed, we take the instability of $Cu^{2+}$ to form $Cu^+$ in the Lead sublattice to support the disproportionation of $2Pb_{10-x}Cu_x(PO_4)_6O$ into $Pb_{20-2x}Cu_{2x}(PO_4)_{12}(O_2)$. A Gedanken synthesis clarifies this further: Consider first synthesizing $Pb_{10-x}(Cu[3d^9])_x(PO_4)_6O$, x=2. Careful heating would allow for $O_2$ evolution to reach the nominal $Pb_8(Cu[3d^{10}])_2(PO_4)_6$ that would in turn resonate with $Pb_8(Cu[3d^9])_2(PO_4)_6$, the former void of "extra" electrons in triangular Lead sites, while the latter analogous to $Pb_{10}(PO_4)_6$, *cf.* Figure 1 again.

5. Finally, preliminary calculations that displace the $Cu^+$ ions to three adjacent phosphate oxygens from 2.04 Å to 1.94 Å, the latter normally corresponding to a $Cu^{2+}$-$O^{2-}$ distance, is found to drastically reduce the gap between the Cu 3d impurity band and the Lead conduction band, telling of phonons possibly driving electron transitions between the two bands.

We refrain from discussing any mechanism for the possible room-temperature superconductivity further at this stage but rather encourage renewed studies on this exciting material, also based on the above. Investigation of the interplay between phonons in the Lead sublattice with copper associated phonons in oxygen deficient $Pb_{10-x}Cu_xO(PO_4)_6$ is encouraged as these jointly control the charge transfer processes between the Lead associated CDW/conduction "flat" bands and the copper associated "ultra-flat" $3d^9$-$3d^{10}$ impurity states. The oxygen deficiency is understood to be increasingly facilitated with increasing x, beyond x=1. Clearly, in ambient air this essential implied oxygen deficiency in LK-99 must be protected.

**Acknowledgement**

Tord Claeson is acknowledged for making me aware of references [1,2] and for showing continuing interest in the hastily evolving process that produced this note. Inspiring discussions with Christine Geers on crystallography are acknowledged.

**Computational details**

The spin-polarized density functional theory calculations utilized the CASTEP program package [6], and the PBESOL GGA implementation therein. The results presented here were found to be insensitive to the choice of U@Cu. P3 symmetry was considered throughout, occasionally checking the results by resorting to P1 symmetry. For further details, please contact the author.